\documentclass[11pt]{amsart}
\usepackage{geometry}                
\usepackage{graphicx}
\usepackage{amssymb}
\usepackage{epstopdf}
\usepackage{colortbl}
\usepackage{tikz}
\def\cs{$\clubsuit$}
\def\ds{$\diamondsuit$}
\def\hs{$\heartsuit$}
\def\ss{$\spadesuit$}
\def\lwidth{0.01cm}
\definecolor{reg_clr}{rgb}{0.84,0.94,0.92}
\definecolor{linear_clr}{rgb}{0.75,0.75,0.75}

\title{Analysis of Solitaire}
\author{Daniel Shiu}

\begin{document}
\maketitle
\section{Introduction}
The Solitaire cipher was designed by Bruce Schneier as a plot point in the novel {\it Cryptonomicon} by Neal Stephenson \cite{cryptonomicon}. The cipher is intended to fit the archetype of a modern stream cipher whilst being implementable by hand using a standard deck of cards with two jokers.

\subsection{Description of Solitaire}
Solitaire is intended to match the design of other modern stream ciphers. In such designs there is an internal state that varies with time $s_t$. There is an update function $U$ that takes states as inputs and returns them as outputs. There is also an extraction function $E$ which takes states as inputs and returns the next part of the key stream. Given an initial state $s_0$, future states are generated by the rule $s_{t+1}=U(s_t)$ and key stream is extracted as $k_t=E(s_t)$ for $t=1,2,\ldots$ A diagrammatic representation is given below.

  \begin{center}
  \hspace{-10pt}{\begin{tikzpicture}
  \draw (0.5,4.5) node[fill=reg_clr,line width=\lwidth,minimum height=0.5cm, minimum width=1cm,draw] (s0) {\textcolor{black}{\small $s_{0}$}};
  \draw (2.5,4.5) node[fill=reg_clr,line width=\lwidth,minimum height=0.5cm, minimum width=1cm,draw] (s1) {\textcolor{black}{\small $s_1$}};
  \draw (4.5,4.5) node[fill=reg_clr,line width=\lwidth,minimum height=0.5cm, minimum width=1cm,draw] (s2) {\textcolor{black}{\small $s_2$}};
  \draw (6.5,4.5) node[fill=reg_clr,line width=\lwidth,minimum height=0.5cm, minimum width=1cm,draw] (s3) {\textcolor{black}{\small $s_3$}};
  \draw (8.5,4.5) node[fill=reg_clr,line width=\lwidth,minimum height=0.5cm, minimum width=1cm,draw] (dots) {\textcolor{black}{\small $\cdots$}};

  \draw[line width=\lwidth] (1.5,4.5) node[fill=linear_clr,minimum height= 0.5cm,circle,inner sep=2pt, draw](u1){\textcolor{black}{\scriptsize $U$}};
  \draw[line width=\lwidth] (3.5,4.5) node[fill=linear_clr,minimum height= 0.5cm,circle,inner sep=2pt, draw](u2){\textcolor{black}{\scriptsize $U$}};
  \draw[line width=\lwidth] (5.5,4.5) node[fill=linear_clr,minimum height= 0.5cm,circle,inner sep=2pt, draw](u3){\textcolor{black}{\scriptsize $U$}};
  \draw[line width=\lwidth] (7.5,4.5) node[fill=linear_clr,minimum height= 0.5cm,circle,inner sep=2pt, draw](u4){\textcolor{black}{\scriptsize $U$}};
  
  \draw[line width=\lwidth] (2.5,2.5) node[fill=linear_clr,minimum height= 0.5cm,circle,inner sep=2pt, draw](e1){\textcolor{black}{\scriptsize $E$}};
  \draw[line width=\lwidth] (4.5,2.5) node[fill=linear_clr,minimum height= 0.5cm,circle,inner sep=2pt, draw](e2){\textcolor{black}{\scriptsize $E$}};
  \draw[line width=\lwidth] (6.5,2.5) node[fill=linear_clr,minimum height= 0.5cm,circle,inner sep=2pt, draw](e3){\textcolor{black}{\scriptsize $E$}};
 
 \draw (2.5,0.5) node[fill=reg_clr,line width=\lwidth,minimum height=0.5cm, minimum width=1cm,draw] (k1) {\textcolor{black}{\small $k_1$}};
 \draw (4.5,0.5) node[fill=reg_clr,line width=\lwidth,minimum height=0.5cm, minimum width=1cm,draw] (k2) {\textcolor{black}{\small $k_2$}};
 \draw (6.5,0.5) node[fill=reg_clr,line width=\lwidth,minimum height=0.5cm, minimum width=1cm,draw] (k3) {\textcolor{black}{\small $k_3$}};

  \draw [-stealth, line width=\lwidth] (s0) -- (u1); 
  \draw [-stealth, line width=\lwidth] (u1) -- (s1);   
  \draw [-stealth, line width=\lwidth] (s1) -- (u2); 
  \draw [-stealth, line width=\lwidth] (u2) -- (s2);   
  \draw [-stealth, line width=\lwidth] (s2) -- (u3); 
  \draw [-stealth, line width=\lwidth] (u3) -- (s3); 
  \draw [-stealth, line width=\lwidth] (s3) -- (u4); 
  \draw [-stealth, line width=\lwidth] (u4) -- (dots); 
  
  \draw [-stealth, line width=\lwidth] (s1) -- (e1); 
  \draw [-stealth, line width=\lwidth] (e1) -- (k1);   
  \draw [-stealth, line width=\lwidth] (s2) -- (e2); 
  \draw [-stealth, line width=\lwidth] (e2) -- (k2);   
  \draw [-stealth, line width=\lwidth] (s3) -- (e3); 
  \draw [-stealth, line width=\lwidth] (e3) -- (k3); 

  \end{tikzpicture}}
  \end{center}
  
 In Solitaire the space of possible states is the space of permutations of 54 elements; equivalently the state consists of the ordering of a deck of cards. This state space is  over 237-bits in logarithmic size which is commensurable with modern stream ciphers. Solitaire can move between a permutation on a deck of 54 cards and the numbers $\{1,2,\ldots,54\}$ by naturally converting the card value to a number between 1-13 and then adding 0, 13, 26, or 39 depending on the suit being \cs, \ds, \hs, or \ss\ (the (distinguished) jokers are taken to be 53 and 54 in this paper, though in the specification both are labelled 53). Using a permutation as state helps to derive a certain uniformity of the key stream: if the sequence of states cannot be distinguished from random permutations and the extraction function samples a single value uniformly from the permutation, then the returned key values should be uniformly distributed across the set which is permuted. Moreover sampling from the permutation can also provide an index into the permutation itself, which again should be uniformly distributed. One needs to be careful however: if the index is sampled from a fixed position $k$ and this value used without modification then this introduces a bias toward $k$ in the extraction (when the value $k$ is in position $k$, it automatically extracts itself; in other circumstances, the extraction will be flat across the space of permutations). Indeed if the extraction functions $S[f(S[k])]$ for any fixed function $f$ and known value $k$, then one should ensure that $f(t)=k$ has no solutions or else a biased stream will be produced. Nevertheless, indexing allows for more complex extraction functions which are uniformly distributed across the permuted set.
 
 This is the approach taken with Solitaire. If we write the state at the time of extraction as  the ordered set $[ S[1], S[2],\ldots,S[54] ]$ where the $S[i]$ take distinct values from the set $\{1,2,\ldots 54\}$, then the extraction function is given by $E([ S[1], S[2],\ldots,S[54] ])=S[S[1]+1]$. A slightly different approach is taken when $S[1]=54$; in this case our indexing exceeds the maximum and instead the key character $S[54]$ is published (in the cipher specification the label is 53 and we reach the same conclusion). This leads to a biased sample position, but should preserve the uniform distribution of the key character rather than introduce a bias towards 54. To put $E$ in the context of our deck of cards, suppose the ordering of the deck from left-to-right is given by:
 
 \medskip
 
\begin{center}\begin{tabular}
{|c|c|c|c|c|c|c|c|c|}
\hline
K\ds & 10\cs & 4\ss & 6\cs & 3\hs & A\cs & 2\hs & jo & A\ds\\
\hline
Q\hs & A\hs & 8\ss & 6\hs & 7\ds & 8\hs & 9\cs &6\ds &4\cs  \\
\hline
5\ds &J\ds &2\cs & K\hs &9\hs &7\hs &Q\ss &7\cs &3\cs \\
\hline
10\hs & 8\cs & 3\ds & JO & 2\ds & J\cs & Q\ds & 8\ds &5\hs \\
\hline
5\ss &K\ss &A\ss & J\ss &10\ss &4\ds &2\ss &9\ds &5\cs \\
\hline
10\ds &3\ss &J\hs & K\cs &9\ss &7\ss &Q\cs &4\hs&6\ss \\
\hline
\end{tabular}\end{center}

\medskip

\noindent This can be converted to the numerical representation and the top card considered numerically:

\medskip

\begin{center} \begin{tabular}
{|c|c|c|c|c|c|c|c|c|}
\hline
26 & 10 & 43 & 6 & 29 & 1 & 28 & 53 & 14\\
\hline
38 & 27 & 47 & 32 & 20 & 34 & 9 & 19 &4  \\
\hline
18 & 24 &2 & 39 & 35 & 33 & 51 &7 &3 \\
\hline
36 & 8 & 16 & 53* & 15 & 11 & 25 & 21 & 31 \\
\hline
44 & 52 & 40 & 50 &49 & 17 & 41 & 22 &5 \\
\hline
23 & 42 & 37 & 13 & 48 & 46 &12 & 30 & 45 \\
\hline
\end{tabular}\end{center}

\medskip

\noindent in this case the number 26 (K\ds). The extracted key value in this case is the 26th card after the top card, which for this deck is 3 (3\cs). The key stream produced is a sequence of numbers in the range $1,\ldots, 54$. This is converted to the more useful key stream taking values $1,\ldots, 26$ by ignoring instances 53 and 54 in the key stream (i.e. jokers) and otherwise reducing modulo 26.

\medskip

\begin{center}\begin{tabular}
{|c|c|c|c|c|c|c|c|c|}
\hline
{\color{red}\bf 26} & 10 & 43 & 6 & 29 & 1 & 28 & 53 & 14\\
\hline
38 & 27 & 47 & 32 & 20 & 34 & 9 & 19 &4  \\
\hline
18 & 24 &2 & 39 & 35 & 33 & 51 &7 &{\color{green}\bf 3} \\
\hline
36 & 8 & 16 & 53* & 15 & 11 & 25 & 21 & 31 \\
\hline
44 & 52 & 40 & 50 &49 & 17 & 41 & 22 &5 \\
\hline
23 & 42 & 37 & 13 & 48 & 46 &12 & 30 & 45 \\
\hline
\end{tabular}\end{center}

The update function $U$ for Solitaire is slightly more complex, though again is intended to be simple enough to implement manually. Firstly the joker corresponding to 53 (hereafter the \emph{slow} joker) is advanced one space in the deck (i.e. for $i$ where $S[i]=53$ apply the swap $(i,i+1)$) and then the joker corresponding to 54 (hereafter the \emph{fast} joker) is advanced two spaces in the deck (i.e. for $i$ where $S[i]=54$ apply the swap $(i,i+1)$ and then the swap $(i+1,i+2)$). If ever this requires us to perform a swap with position 55 (i.e. the joker is the last card of the deck) then instead of a swap perform a single right rotation of the last 53 cards or in cycle notation we apply $(2,3,\ldots,53,1)$ (i.e. the joker is moved to the position $S[2]$ and all subsequent cards increase their index by 1). Here is an illustration of the joker movement for our example:

  \begin{center}\begin{tabular}
{|c|c|c|c|c|c|c|c|c|}
\hline
26 & 10 & 43 & 6 & 29 & 1 & 28 & {\color{red}\bf 53} & 14\\
\hline
38 & 27 & 47 & 32 & 20 & 34 & 9 & 19 &4  \\
\hline
18 & 24 &2 & 39 & 35 & 33 & 51 &7 &3 \\
\hline
36 & 8 & 16 & {\color{red}\bf 54} & 15 & 11 & 25 & 21 & 31 \\
\hline
44 & 52 & 40 & 50 &49 & 17 & 41 & 22 &5 \\
\hline
23 & 42 & 37 & 13 & 48 & 46 &12 & 30 & 45 \\
\hline
\end{tabular}\end{center}

  \begin{center}\begin{tabular}
{|c|c|c|c|c|c|c|c|c|}
\hline
26 & 10 & 43 & 6 & 29 & 1 & 28 & {\color{green}\bf 14} & {\color{red}\bf 53}\\
\hline
38 & 27 & 47 & 32 & 20 & 34 & 9 & 19 &4  \\
\hline
18 & 24 &2 & 39 & 35 & 33 & 51 &7 &3 \\
\hline
36 & 8 & 16 & {\color{green} \bf 15} & {\color{green}\bf11} & {\color{red}\bf 54} & 25 & 21 & 31 \\
\hline
44 & 52 & 40 & 50 &49 & 17 & 41 & 22 &5 \\
\hline
23 & 42 & 37 & 13 & 48 & 46 &12 & 30 & 45 \\
\hline
\end{tabular}\end{center}
  
\medskip

The next stage of $U$ is to perform a ``triple cut''. Here all cards prior to the first joker are moved to the end of the deck and all cards after the second joker are moved to the beginning. In other words if the first joker occupies position $i$ and the second joker position $j$ with $i>j$ we apply the permutation 
$$j+1, j+2,\ldots, 54, i, i+1,\ldots, j, 1,2,\ldots, i-1$$ 
(note that this is not in cycle notation). Again in our example here is an illustration of a triple cut:

\medskip

  \begin{center}\begin{tabular}
{|c|c|c|c|c|c|c|c|c|}
\hline
26 & 10 & 43 & 6 & 29 & 1 & 28 & 14 & {\color{red}\bf 53}\\
\hline
38 & 27 & 47 & 32 & 20 & 34 & 9 & 19 &4  \\
\hline
18 & 24 &2 & 39 & 35 & 33 & 51 &7 &3 \\
\hline
36 & 8 & 16 & 15 & 11 & {\color{red}\bf 54} & 25 & 21 & 31 \\
\hline
44 & 52 & 40 & 50 &49 & 17 & 41 & 22 &5 \\
\hline
23 & 42 & 37 & 13 & 48 & 46 &12 & 30 & 45 \\
\hline
\end{tabular}\end{center}

  \begin{center}\begin{tabular}
{|c|c|c|c|c|c|c|c|c|}
\hline
 25 & 21 & 31 & 44 & 52 & 40 & 50 &49 & 17 \\
 \hline
 41 & 22 & 5 & 23 & 42 & 37 & 13 & 48 & 46\\
 \hline 
 12 & 30 & 45 & {\color{red}\bf 53} & 38 & 27 & 47 & 32 & 20 \\
 \hline
  34 & 9 & 19 & 4 & 18 & 24 & 2 & 39 & 35\\
  \hline 
  33 & 51 &7 &3 & 36 & 8 & 16 & 15 & 11\\
  \hline
  {\color{red}\bf 54} & 26 & 10 & 43 & 6 & 29 & 1 & 28 & 14\\
   \hline
\end{tabular}\end{center}

\medskip

Finally we apply a ``count cut''. Here we look at the last card of the deck $S[54]$ whose values we denote $k$. We then cut the first 53 cards so that the $(k+1)$th card $S[k+1]$ becomes the top card $S[1]$. In permutation notation we apply the permutation 
$$k+1,k+2\ldots 53,1,2,\ldots k,54$$ 
(again note that this is not cycle notation). Similar to the extraction function, Solitaire introduces a slight modification when $S[54]=54$. In this case no modification is made during the count cut step (in fact nor is any modification is made if $S[54]=53$). Completing our diagrams here is an example of the count cut:

\medskip

  \begin{center}\begin{tabular}
{|c|c|c|c|c|c|c|c|c|}
\hline
 25 & 21 & 31 & 44 & 52 & 40 & 50 &49 & 17 \\
 \hline
 41 & 22 & 5 & 23 & {\color{green}\bf 42} & 37 & 13 & 48 & 46\\
 \hline 
 12 & 30 & 45 & 53 & 38 & 27 & 47 & 32 & 20 \\
 \hline
  34 & 9 & 19 & 4 & 18 & 24 & 2 & 39 & 35\\
  \hline 
  33 & 51 &7 &3 & 36 & 8 & 16 & 15 & 11\\
  \hline
  54 & 26 & 10 & 43 & 6 & 29 & 1 & 28 & {\color{red}\bf14}\\
   \hline
\end{tabular}\end{center}  

  \begin{center}\begin{tabular}
{|c|c|c|c|c|c|c|c|c|}
\hline
37 & 13 & 48 & 46 & 12 & 30 & 45 & 53 &38\\
\hline 
27 & 47 & 32 & 20 & 34 & 9 & 19 & 4 &18\\
\hline
24 & 2 & 39 & 35 & 33 & 51 &7 & 3 &36\\
\hline
8 & 16 & 15 & 11 & 53* & 26 & 10 & 43 & 6\\ 
\hline 
29 & 1 & 28 & 25 & 21 & 31 & 44 & 52 & 40  \\
\hline 
50 &49 & 17 & 41 & 22 & 5 & 23 &  {\color{green}\bf 42} & {\color{red}\bf14}\\
 \hline
\end{tabular}\end{center}
  
\section{Empirical observations}
It is easy to empirically test the statistical properties of the key stream for lengths of $2^{32}$ characters. Freely available C implementations can generate this much key stream in a few minutes on a laptop. Combining this with a randomly shuffled deck (based on entropy for {\tt /dev/random}) it is easy to confirm that the keystream is indeed uniformly distributed across individual characters. However, as reported elsewhere \cite{crowley}, there is an empirically observable bias in the difference between consecutive key characters. It is again easy to confirm that consecutive characters mod 26 are equal with probability roughly $0.0444\approx 1/22.5$. One can modify the program to directly output the keystream mod 54 and observe that the bias towards equality here is about $0.0254\approx 1/39.5$ which suggests that this is due to repeating keycards rather than any interplay between values that differ by 26.

One can investigate repetitions at other short distances within the key stream and although not as pronounced, empirical methods can detect a bias for characters even separated by as much as 26 characters. The exact bias at the various separations is plotted below.

\begin{center}\includegraphics[height=6cm]{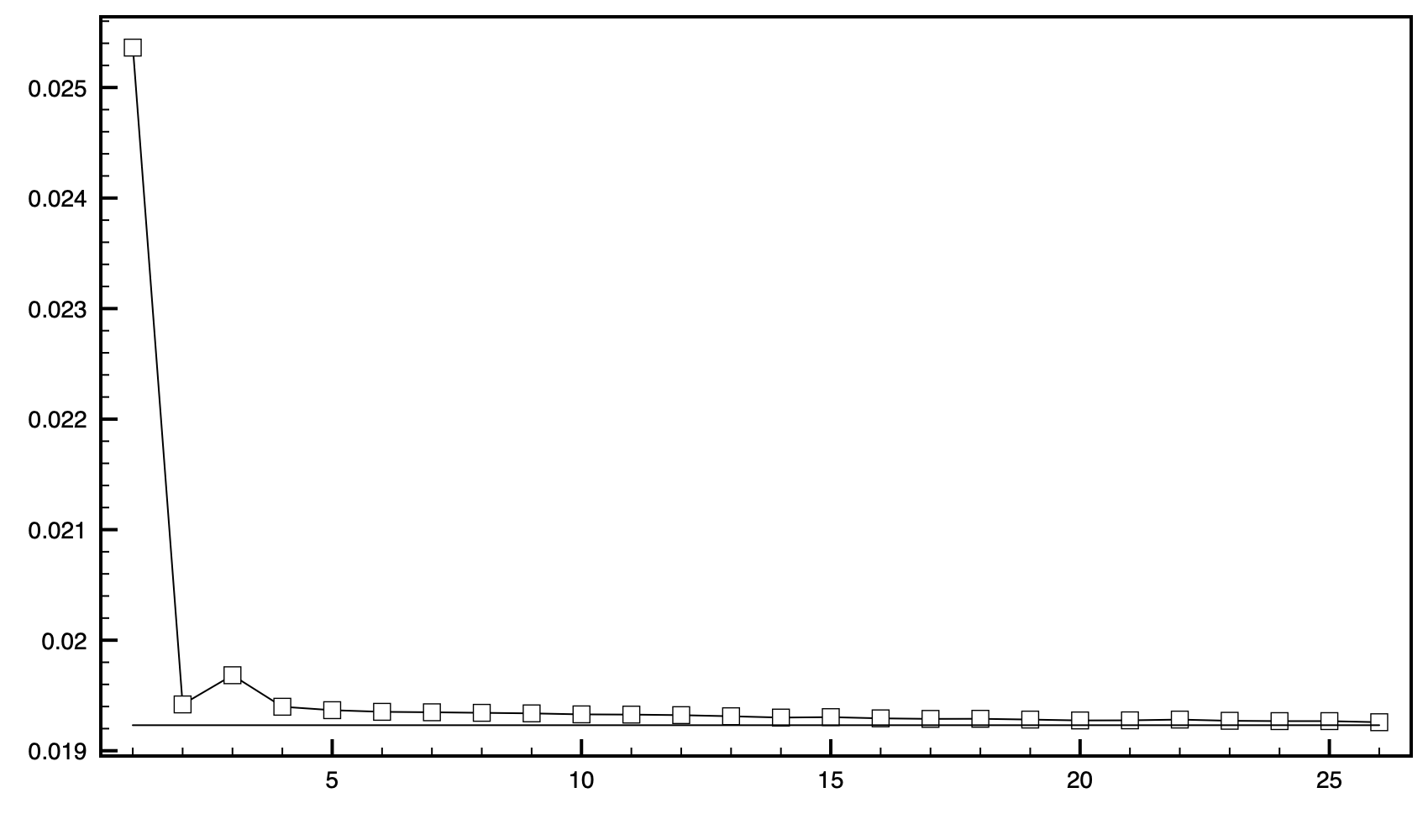}\label{diag:freq-delta}\end{center}

One can also count occurrences of repeated triples, four tuples and five tuples. These occur with probability roughy $4.89\times 10^{-4}$, $1.36\times 10^{-5}$ and $6.81\times 10^{-6}$ respectively. Note that if we take the ratio of the first term with 0.0444 and the ratios of subsequent terms we get 90.8, 36.0, and 2.0 rather than 22.5 (which we might expect one each case if occurrences of the bias between consecutive terms were an independent event). Indeed, the occurrences of triples and four-tuples are \emph{rarer} than for a random stream.

\subsection{Quantifying the threat of the main bias}
One can quantify the risk of the most significant bias. Using Shannon's entropy function $H$ we consider a stream of values mod 26 where after the initial character, the next character in the stream is identical to the previous with probability $\approx 1/22.5$ or is selected uniformly at random from the remaining 25 characters with probability $\approx 21.5/22.5$. The entropy of each character after the first in such a stream is given by
	$$-\frac1{22.5}\log\frac1{22.5}-25\cdot\frac{21.5}{22.5}\cdot\frac1{25}\log\frac{21.5}{22.5\cdot25}$$
which is roughly 3.2577 if measured in the natural logarithm or 4.6999 if measured in bits. These values should be compared to the entropy of a flat character stream mod 26 where each character is selected independently and uniformly at random which is $\log 26$ (roughly 3.2580 in the natural logarithm or 4.7004 in bits). Thus using only the information from the principal bias we conclude that Solitaire leaks information at a rate of 0.0005 bits per character produced after the first.

Response to this rate of information loss will vary. The theorist cryptographer will hold that any rate of information loss is unacceptable. A more pragmatic-minded cryptographer might reflect that well-established methods of creating a stream cipher, such as a block cipher used in CTR or OFB mode leak information at a rate of $2^{-B-1}t^2$ after producing $t$ blocks of $B$ bits of key stream (at least for values of $t\ll 2^B$). For short messages, at most hundreds of characters long, this bias does not seem too harmful. For longer messages we offer a couple of cautionary examples.

Suppose that Solitaire were used to secure logon credentials so that users repeatedly send a user name and password encrypted under Solitaire (possibly re-keyed between sessions). Suppose that the credentials are a total of 30 characters long, but are sent over 50,000 times. Counting the occurrences of all possible differences between the first and second cipher text character in the set of 50,000 we expect difference to occur about 2222 times owing to our bias. We expect other differences to be binomially distributed with mean about 1911 and standard deviation about 42 so that a random difference occurring more often than our bias is unlikely. Repeating the process for all the other differences gives us very strong guesses for the difference between consecutive characters in the plaintext and allows us to limit the possible user name and password to one of twenty-six choices which are Caesar shifts of each other.

As another example, suppose a suspected dissident is accused of leaking a government document using Solitaire and that the document is 10,000 characters. In their defence the suspect produces an innocuous document of the same length. A cryptographic expert witness could derive the key stream under each claim and count the repeats. The causal case would be expected to have about 444 repeats in the key stream and the non-causal around 385. The difference of 60 would represent more than 2 standard deviations, which may constitute reasonable doubt dependent on the legal system and the member of the judiciary\begin{footnote}{Note for lawyers: it would be simple enough to construct a plausible message that leads to a causal-looking key stream for any given cipher texts and so both parties would need to demonstrate that their claimed message was produced independently of the cipher text.}\end{footnote}.

\section{Explaining the main bias}
A bias to repetition in the key stream is likely caused by a similarity of two consecutive states causing the same extraction for the same reason. Here is one example of how such a repetition might occur. Our story depends principally on the value of the top card, the locations of the jokers and the value of the card which will become the bottom card after our update. Consider the pack below where a key extraction has just occurred.The top card, 3, causes the key card 6 to be indexed and published.

\medskip 

  \begin{center}{\begin{tabular}
{|c|c|c|c|c|c|c|c|c|}
\hline
{\color{red}\bf 3} & 10 & 43 & {\color{green}\bf 6} & 29 & 1 & 28 & 53 & {\color{blue}\bf 46}\\
\hline
38 & 27 & 14 & 32 & 20 & 34 & 9 & 19 &4  \\
\hline
18 & 24 &2 & 39 & 35 & 33 & 51 &7 & 26 \\
\hline
36 & 8 & 16 & 54 & 15 & 11 & 25 & 21 & 31 \\
\hline
44 & 52 & 40 & 50 &49 & 17 & 41 & 22 &5 \\
\hline
23 & 42 & 37 & 13 & 48 & 47 &12 & 30 & 45 \\
\hline
\end{tabular}}\end{center}

\medskip

We now proceed to apply $U$ by advancing the jokers, noting that the card to the left of the first joker, 46, is designed to become our bottom card.

\medskip

  \begin{center}{\begin{tabular}
{|c|c|c|c|c|c|c|c|c|}
\hline
{\color{red}\bf 3} & 10 & 43 & {\color{green}\bf 6} & 29 & 1 & 28 &  {\color{blue}\bf 46} & 53\\
\hline
38 & 27 & 14 & 32 & 20 & 34 & 9 & 19 &4  \\
\hline
18 & 24 &2 & 39 & 35 & 33 & 51 &7 & 26 \\
\hline
36 & 8 & 16 & 15 & 11 & 54 & 25 & 21 & 31 \\
\hline
44 & 52 & 40 & 50 &49 & 17 & 41 & 22 &5 \\
\hline
23 & 42 & 37 & 13 & 48 & 47 &12 & 30 & 45 \\
\hline
\end{tabular}}\end{center}

\medskip

We now perform a triple cut which moves 46 to the bottom. We note that the order of all cards between 3 and 46 is preserved.

\medskip

  \begin{center}{\begin{tabular}
{|c|c|c|c|c|c|c|c|c|}
\hline
 25 & 21 & 31 & 44 & 52 & 40 & 50 &49 & 17\\
\hline
 41 & 22 &5 & 23 & 42 & 37 & 13 & 48 & 47\\
\hline
12 & 30 & 45 & 53 & 38 & 27 & 14 & 32 & 20\\
\hline
 34 & 9 & 19 & 4 & 18 & 24 & 2 & 39 & 35 \\
\hline
 33 & 51 &7 & 26 & 36 & 8 & 16 & 15 & 11\\
\hline
 54 & {\color{red}\bf 3} & 10 & 43 & {\color{green}\bf 6} & 29 & 1 & 28 &  {\color{blue}\bf 46} \\
\hline
\end{tabular}}\end{center}

\medskip

We now perform a count cut. Due to the particular value of 46, this restores 3 to the top card position as well as the subsequent cards up to 46.

\medskip

\begin{center}{\begin{tabular}
{|c|c|c|c|c|c|c|c|c|}
\hline
 {\color{red}\bf 3} & 10 & 43 & {\color{green}\bf 6} & 29 & 1 & 28 & 25 & 21\\
 \hline  
 31 & 44 & 52 & 40 & 50 &49 & 17& 41 & 22\\ 
 \hline
 5 & 23 & 42 & 37 & 13 & 48 & 47 & 12 & 30\\
\hline 
45 & 53 &38 & 27 & 14 & 32 & 20 & 34 & 9\\
\hline
 19 & 4 & 18 & 24 &2 & 39 & 35 & 33 & 51\\
 \hline
 7 & 26 & 36 & 8 & 16 & 15 & 11 & 54 &   {\color{blue}\bf 46} \\
\hline
\end{tabular}}\end{center}

If we now perform a second extraction we do of course recover the value 6 again.

We now consider the probability that this ``story'' occurs in a more generic fashion. The conditions depend on four cards:
\begin{itemize}
\item The card at the top of the deck when the extraction is first made, call this the dereferenced card (3 in our example),
\item The card that moves to the bottom of the deck during the update, call this the count card (46 in our example),
\item The two jokers (53 and 54 in our example).
\end{itemize}
In the case where the count card is passed by the slow joker we enumerate the conditions for the $54\times53\times52\times51$ possible values of these cards that lead to such a phenomenon. We consider the case where the slow joker is in position $i$ before advancing and position $i+1$ afterwards. After moving jokers we require the count card to be in position $i$ and to take the value $55-i$. Now, if the fast joker is in a position greater than $i+1$ after joker movement (are $53-i$ such positions), our dereferenced card will be restored to the top of the deck, along with the $i-2$ subsequent cards from the previous state. Thus if the derefenced card is $i-2$ or less in value, the same key card is extracted. If $i<29$ all values between 1 and $i-2$ could potentially be assumed by the dereferenced card, but for $i\ge 29$ only $i-3$ values are available as the dereferenced card cannot be $55-i$. For the case where the slow joker passes the count card then, the probability of this phenomenon is
   $$\frac1{54\cdot53\cdot52\cdot51}\left(\sum_{i=3}^{28}1\cdot1\cdot(53-i)\cdot(i-2)+\sum_{i=29}^{52}1\cdot1\cdot(53-i)\cdot(i-3)\right)=\frac{21800}{54\cdot53\cdot52\cdot51}.$$
 Reasoning similarly for the fast joker there is another contribution
    $$\frac1{54\cdot53\cdot52\cdot51}\left(\sum_{i=3}^{28}1\cdot1\cdot(53-i)\cdot(i-3)+\sum_{i=29}^{52}1\cdot1\cdot(53-i)\cdot(i-4)\right)=\frac{20525}{54\cdot53\cdot52\cdot51}.$$
Our heuristic tells us to expect such an arrangement to cause a repeat in the key stream roughly $42325/7590024\approx 0.0055764$ of the time. If we denote this probability $p$ and assume that outside this scenario the chance of a repeat is 1/26, then we should expect to observe a repeat rate of
   $$p+\frac{(1-p)}{26}\approx 0.043823$$
which is indeed close to the observed repeat rate of 0.444. This would appear to explain most of the bias and is borne out when we count the occurrences of the top card of the deck on occasions when the extracted card is a repeat of the previous extraction:

\begin{center}\includegraphics[height=6cm]{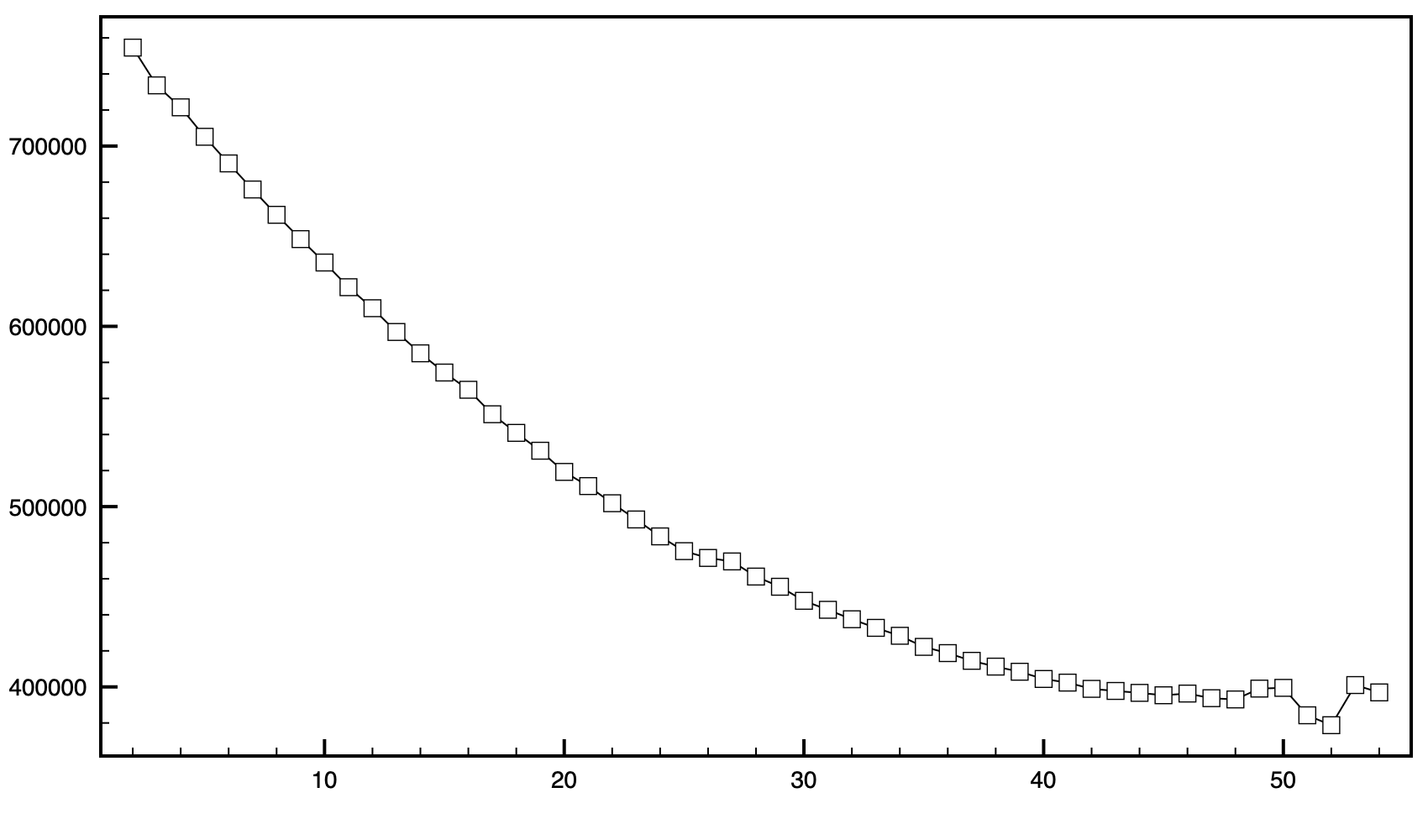}\label{diag:top_repeat}\end{center}

As our model predicts, repeats are most frequent when the dereferenced card is a small value. Note also the slight kink in the graph between 28 and 29 corresponding to the separation in our sums. The tail of this graph suggests an additional phenomenon around jokers. It follows that there are likely to be other sporadic causal constructions.

One might wonder why this situation has such reduced effect if we look at extractions 2 time steps apart, or why it fails to propagate to longer runs of three, four or five repeats. This is because the top card always ends directly after a joker following the triple cut. If these are not then separated by the count cut, the joker will pass the former top card at the next update and change the distance between the former top card and the extracted card. Even if the top card is restored by the count cut, the joker will be in position 53. If this joker is the fast joker, again the distance will be disrupted on the next update.

\section{Modifications to reduce the bias}
The story of the previous section explains how this particular design leads to a bias towards repeats in the keystream. The more general design issue here is combination of two effects:
\begin{itemize}
\item The update function $U$ does not mix thoroughly,
\item The extraction function $E$ does not ``avalanche'' well.
\end{itemize}
By ``avalanche'' we mean the general cryptographic property that all small changes to the input of a cryptographic function should be enough to render the output indistinguishable from random. We note that it does not seem to be necessary that neither of these flaws occur in a generic secure stream cipher. For example, a block cipher in CTR mode has a very simple update function (incrementing a counter) and a block cipher in OFB mode has a very simple extraction function (the identity); in both cases however the partner function is (assuming that the block cipher design is well-designed) very strong indeed.

One can quantify the mixing $U$ by using one of the many measures of the complexity of a permutation. These include \emph{inter alia} counting ascents, counting ascending runs, and counting inversions. The issues within Solitaire however seem to be caused by $U$ preserving long stretches of the state (though possibly translating these). A sensible complexity measure would therefore seem to be to count adjacencies. For a permutation $\pi$ on $n$ elements the number of adjacencies is the size of the set $\{1\le i\le n-1: \pi_{i+1}=\pi_i+1\}$. The expected number of adjacencies for a random permutation is 1, but if a permutation preserves a stretch $r$ elements then the number of adjacencies is at least $r-1$.

We consider the typical number of adjacencies changed by application of $U$. Advancing a joker typically leaves 50 adjacencies intact \begin{footnote}{If the joker is at the bottom of the deck, only two adjacencies are changed. The two joker advances combined will typically leave 47 adjacencies intact, but possibly fewer.}\end{footnote}. The triple cut leaves at least 51 adjacencies intact, but the adjacencies removed were already affected by the joker movement. The count cut leaves 51 adjacencies intact, there is some possible overlap with previous effects. The overall application of $U$ is therefore expected to preserve at least 45 adjacencies rather than the expected 1 preserved by a ``random'' permutation. By this measure then, the function $U$ falls along way short of the mixing desired.

One could attempt to increase the level of mixing by applying the given update function multiple times. The above analysis shows that at least 7 applications of $U$ will be necessary to expect at most one adjacency to be preserved.  Although the adjacencies affected by repeated application of $U$ are probably not independent (consider the adjacencies affected by joker movement), one can treat them as such. Following this line of analysis, one might model the expected number of adjacencies preserved by $s$ applications of $U$ as $53(45/53)^s$ until this expectation drops below 1 (here we ignore the effect of an application restoring an adjacency from the original state). This suggest that 25 applications of $U$ should be considered, an estimate approximately born out by diagram \ref{diag:freq-delta}. However, applying the $U$ function 25 times between key extraction probably violates the desired design feature that the cipher be easy to implement by hand.

The alternative is to modify the extraction function to avalanche better. At present the $E$ function depends on the value of only two cards and if these two cards maintain their position then the extraction function behaves in the same way. A more complex extraction function might involve more indexing. One might consider using the values of, say,  both $S[1]$ and $S[2]$ to generate an index from which to extract e.g. use the extraction function $S[S[1]+S[2]]$. However, the chance of preserving both $S[1]$ and $S[2]$ is not much greater than the chance of preserving just $S[1]$. In this case rather than repeats occurring from a small value of $S[1]$ repeats will occur from a small value of $S[1]+S[2]\pmod{54}$ which is almost equiprobable (the distribution of $S[1]+S[2]\pmod{54}$ is slightly biased towards odd values).

One could extract by adding another layer of dereferencing e.g. $S[S[S[1]+1]+1]$. This does avalanche with the update much better. The sums from section 3 gain an additional factor of $(i-3)/50$, $(i-4)/50$ or $(i-5)/50$. Further layers of dereferencing have a similar effect. Again the reduction in bias comes with an undesirable overhead for the user. One also needs to be careful as iterated dereferencing can coalesce when $S[54]=53$ or $S[53]=54$ by the extraction rules of Solitaire (though the deletion of these key values will ensure that the key stream is not polluted). Additional factors $(54-i)/55$ could be applied to the sums in section 3 if we add another joker, but again the reduction in bias comes at a cost to the user.

Probably the best way to proceed is to lightly modify both the update and extraction functions (making both of these of moderate complexity would appear to be a good principle of stream cipher design). The question is how much complexity can be added before over-burdening the user.

\section{Cycle structure of $U$}
Another objection raised by \cite{crowley} is that the update function $U$ in Solitaire is not bijective due to the exceptional joker movement when the end of the pack is reached. The argument is that the cycle structure of a random bijective map provides a greater likelihood of being on a long cycle (the most likely outcome is to land on a cycle of length $\sim (1-1/e)n$ for a state space of size $n$) than a random map (where the most likely outcome is to coalesce to a cycle of length $O(n^{1/2})$ after $O(n^{1/2})$ steps). Short cycles lead to instances of repeating key stream which is highly insecure.

In the case of Solitaire however the cycle structure of $U$ is unlikely to be as bad as a completely random map. The image of $U$ is still very large (i.e. there are very few states which cannot be reached by applying $U$ to a different state) and exact counts can be made of states with 0, 1, 2, and 3 pre-images (higher numbers of pre-images are not possible). Analysis of random functions with particular configurations of pre-images can be performed with the configuration model \cite{config}. Under this model, the transition to the random map ``giant component'' will happen when a significant proportion of states have at least 2 pre-images. The $U$ function is a long way from this condition and we should expect it to behave very similarly to a random bijection.

Moreover,  a non-bijective $U$ means that if an internal state is recovered then earlier states cannot necessarily be uniquely recovered and a modicum of forward security is provided.

\section{Conclusion}
We have found a model for repetitions in the keystream in the stream cipher Solitaire that accounts for the large majority of the repetition bias. Other phenomena merit further investigation. We have proposed modifications to the cipher that would reduce the repetition bias, but at the cost of increasing the complexity of the cipher (probably beyond the goal of allowing manual implementation). We have argued that the state update function is unlikely to lead to cycles significantly shorter than those of a random bijection.


\begin{thebibliography}{References}

\bibitem{crowley} P. Crowley, \emph{Problems with Bruce Schneier's "Solitaire"\hfil}, \texttt{http://www.ciphergoth.org/crypto/solitaire/}
\bibitem{config} M. Molloy and B. Reed, \emph{A critical point for random graphs with a given degree sequence}, Random Structures \& Algorithms. \textbf{6} (2-3) (1995) pp. 161Ð180.
\bibitem{cryptonomicon} N. Stephenson, \emph{Cryptonomicon}, Avon, 1999. ISBN 0-380-97346-4


\end{thebibliography}
\end{document}